\documentclass[twocolumn,showpacs,amsmath,amssymb,prl]{revtex4}

\usepackage{graphicx}
\usepackage{dcolumn}
\usepackage{bm}
\usepackage{epsf}
\usepackage{amsmath}
\usepackage{amssymb}

\newcommand{\tr}[1]{\text{tr}\left\{#1\right\}}

\newcommand{\la}{\left\langle}
\newcommand{\ra}{\right\rangle}
\newcommand{\pd}{\partial}
\newcommand{\bla}{bla\\bla\\bla\\bla\\}
\newcommand{\PRA}{Phys. Rev. A }

\newcommand{\PRE}{Phys. Rev. E }
\newcommand{\PRL}{Phys. Rev. Lett. }

\newcommand{\e}[1]{\exp{\left( #1 \right)}}

\begin{document}

\title{Nonequilibrium entropy production for  open quantum systems}
\author{Sebastian Deffner and Eric Lutz}
\affiliation{Department of Physics, University of Augsburg, D-86135 Augsburg, Germany}

\begin{abstract}
We consider open quantum systems weakly coupled to a heat reservoir and driven by  arbitrary time-dependent parameters. We derive exact microscopic expressions for the nonequilibrium entropy production and entropy production rate, valid arbitrarily far from equilibrium. By using the two-point energy measurement statistics for system and reservoir, we further obtain a quantum generalization of the integrated fluctuation theorem put forward by Seifert [PRL 95, 040602 (2005)]. \end{abstract}

\maketitle

Nonequilibrium phenomena are ubiquitous in nature. Yet, a general framework allowing their description far from equilibrium is lacking \cite{leb08}. 
 A defining feature of out-of-equilibrium  systems is that they dissipate energy in the form of heat, leading to an irreversible increase of their entropy. The nonequilibrium entropy production $\Sigma$ and its time derivative, the entropy production rate $\sigma=\pd_t \Sigma$, are therefore two fundamental concepts of nonequilibrium thermodynamics \cite{leb08}. Traditionally, the entropy production is used to evaluate the performance of thermodynamic devices: the maximal useful work, the  availability (or exergy),  that can be extracted  from a given system, is reduced by the presence of  irreversibilities, such as friction or nonstatic transformations. This  loss  of availability is directly related to the mean entropy production $\la\Sigma\ra $ \cite{cen01}. On the other hand, the entropy production rate is the generating function of the nonequilibrium fluxes. Consider  a system driven away from equilibrium by some thermodynamic forces $X_i$, for example an electric field or a temperature difference. These forces will cause unknown nonequilibrium fluxes $J_i$, for instance an electric current or a heat flow, that are given by the derivative of the entropy production rate with respect to the applied force, $J_i= \pd \sigma/\pd X_i$ \cite{gal99}.    More recently, the entropy production has been instrumental in the analysis of  nonequilibrium effects in many different branches of physics \cite{tie06,tur07,and10}, including the study of quantum impurity models \cite{meh08} and of quenched quantum many-body systems \cite{pol11}. The explicit expression of the mean quantum entropy production  $\la \Sigma\ra$ is in general unknown, however.  A remarkable property of the entropy production  is that it satisfies a fluctuation theorem of the form  $\la \exp(-\Sigma)\ra = 1$, which holds arbitrarily far from equilibrium \cite{eva93,gal95}. A  unified version of   classical fluctuation theorems, valid for nonequilibrium initial conditions and arbitrary driving,   has lately  been obtained by Seifert for the total entropy change occurring in system and  reservoir \cite{sei05}.

In this paper, we provide generic microscopic expressions for the entropy production, and the entropy production rate, for open quantum systems that are weakly coupled to a heat reservoir  and  driven arbitrarily far from equilibrium by external parameters.  The time evolution and the thermodynamic properties of weakly damped quantum systems are usually  described by Markovian master  equations of the Lindblad type \cite{ali79,lin83,bre02}. With their help,  microscopic expressions for the entropy production rate in terms of the reduced density  operator of the system have been obtained for relaxation \cite{spo78}, transport \cite{spo78a} and slowly driven quantum processes \cite{bre03}. Moreover,  a fluctuation theorem for the entropy produced along quantum trajectories has been derived in Refs.~\cite{roe04,esp06,roe08}. However, Markovian master equations are limited to slow driving, that is near equilibrium processes, as their derivation is based on the assumption of time-independent Hamiltonians \cite{dav78}. A formulation of  completely positive maps for fast driven quantum systems, that is far from equilibrium transformations, for which the Markovian approximation is likely to break down, appears difficult \cite{len86,bre04}. In the following, we employ a  thermodynamic approach to derive the exact mean nonequilibrium quantum entropy production and mean quantum entropy production rate without relying on master equations or on quantum trajectories. The obtained expressions are therefore valid for driving processes that operate arbitrarily far from equilibrium. In addition,  starting from the two-point energy measurement statistics for system and reservoir, we derive a quantum extension of Seifert's fluctuation theorem \cite{sei05}. Our general formalism allows us  to recover and extend a number of previously known results in a \mbox{unified manner}.

\textit{Nonequilibrium quantum entropy production.} We consider a quantum system whose Hamiltonian $H_t$ is driven by an arbitrary time-dependent external parameter during a finite time interval $\tau$. The system is assumed to be weakly coupled to an infinitely large  thermal reservoir with which it can exchange energy in the form of heat \cite{rem}. We  focus for the time being on  the situation where the system is initially and finally in an equilibrium  (or at least local equilibrium) state, so that its thermodynamic variables---in particular its entropy---are well defined. We emphasize that the quantum system need not be in equilibrium {\it with} the reservoir at these times (it can for instance be at another temperature), nor does it need to remain close to equilibrium during the driving process (see Fig. 1). We denote by $\Delta U = U_\tau-U_0$ the change of internal energy between final and initial states and by $\Delta S= S_\tau-S_0$ the corresponding change  of entropy. Since $U$ and $S$ are state functions, their variations are process independent. According to the laws of  thermodynamics, the changes of internal energy and entropy are  \cite{leb08}
\begin{eqnarray}
 \Delta U &=& \la W\ra + \la Q\ra  \ , \label{1}\\
\Delta S &=& \beta \la Q\ra + \la \Sigma \ra  \label{2}\ ,
\end{eqnarray}
where $\la W\ra $ is the mean work done on the system, $\la Q\ra$ the mean heat exchanged during the process, and $\la \Sigma\ra$ the mean entropy production ($\beta=1/T$ is the inverse temperature of the reservoir).  It is worth noticing that these quantities are well-defined for arbitrary nonequilibrium transformations. In particular, the thermodynamic entropy may not be defined out of equilibrium, but the entropy production between initial and final equilibrium states always is. From Eqs.~(1) and (2) we see that the mean nonequilibrium entropy production is given by,
\begin{equation}
\label{3}
\la \Sigma \ra= \Delta S -\beta \Delta U +\beta \la W \ra \ .
\end{equation}
In order to obtain a microscopic expression for $\la \Sigma\ra$, we introduce the reduced density operator $\rho_t$ of the system.  The initial and final equilibrium  operators are $\rho_i= \exp(-\beta_i H_0)/Z_i$ and  $\rho_f= \exp(-\beta_f H_\tau)/Z_f$, where $Z_i$ and $Z_f$ are the partition functions. We further define the instantaneous equilibrium operator $\rho^\text{eq}_t = \exp(-\beta H_t)/Z_t$, with respect to the reservoir, that corresponds to the Hamiltonian $H_t$. The initial entropy is then of the form,
\begin{equation}
\label{4}
 S_i=-\tr{\rho_i  \ln{\rho_i}}\ .
\end{equation}
At the same time,  the initial  internal energy is,
\begin{equation}
\label{5}
\beta U_i= \beta\tr{\rho_i H_{0}}=-\tr{\rho_i\ln{\rho_0^\text{eq}}}+\ln{Z_0}\ ,
\end{equation}
where we have used $Z_0=\tr{\exp{\left(-\beta H_0\right)}}$. Similar expressions are obtained for the final entropy and internal energy by replacing $i$ by $f$ and $0$ by $\tau$. We identify the mean work done during the process  with the average change of the Hamilton with time \cite{ali79},
\begin{equation}
\begin{split}
\label{6}
 \beta \la W\ra&=\beta \int_0^\tau dt\,\tr{\rho_t\,\pd_t H_t}\\&=-\int_0^\tau dt\, \tr{\rho_t \,\pd_t \ln{\rho^\text{eq}_t}}-\ln{Z_\tau}+\ln{Z_0} \ .
\end{split}
\end{equation}
Combining Eqs.~\eqref{3}-\eqref{6}, we obtain the general expression, 
\begin{equation}
\label{7}
 \la \Sigma\ra=S\left(\rho_i||\rho_0^\text{eq}\right)-S\left(\rho_f||\rho^\text{eq}_\tau\right)-\int_0^\tau dt\,\tr{\rho_t\, \pd_t \ln{\rho}_t^\text{eq}}\ ,
\end{equation}
where we have introduced the quantum relative entropy of two operators  
$ S\left(\rho_1||\rho_2\right)=\tr{\rho_1\ln{\rho_1}-\rho_1\ln{\rho_2}}$ \cite{nie00}. Equation \eqref{7} is the exact microscopic expression for the mean  nonequilibrium entropy production for a driven  open quantum system weakly coupled to a single heat reservoir. It is valid for intermediate states that can be arbitrarily far from equilibrium.  We next focus on four special cases  to elucidate the physical meaning of the different terms appearing in Eq.~\eqref{7}, 

\begin{figure}
\includegraphics[width=0.44\textwidth,angle=0]{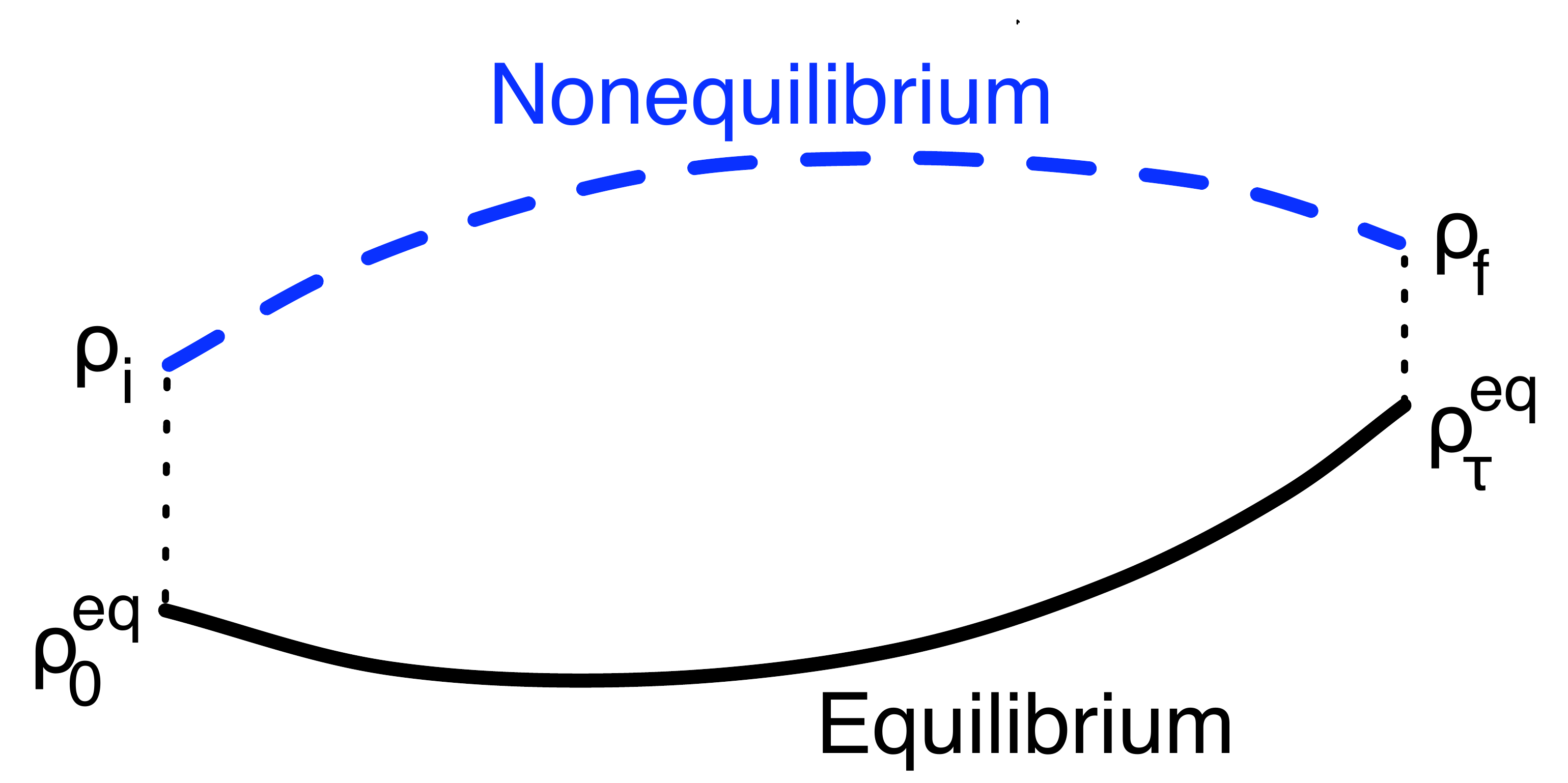}
\caption{ During a quasistatic process,  a weakly damped system remains in equilibrium with the reservoir at all times. We here consider open quantum systems that are driven by arbitrary time-dependent external  parameters and may initially and finally be  in any  state $\rho_i$ and $\rho_f$, not necessarily in equilibrium with the reservoir.  \label{fig1}}
\end{figure}

a. Let us start with the case of an undriven system, $\pd_t H_t=0$, which corresponds to a  pure relaxation process (without work). Since $\rho_f=   \rho_\tau^\text{eq}$, Eq.~\eqref{7} \mbox{reduces to}  
\begin{equation}
\label{9}
 \la \Sigma\ra_\text{equil.}=S\left(\rho_i||\rho_0^\text{eq}\right) \ .
\end{equation}
This is the mean entropy production associated with the equilibration of the quantum
system with the reservoir, a result first obtained by Schl\"ogl for classical nonequilibrium systems \cite{sch66} (see also Refs.~\cite{pro76,sch80}).

b. Assume the  system is  initially in equilibrium with the reservoir, $\rho_i = \rho_0^\text{eq}$ (no initial equilibration). When the driving is such that the system also ends in  equilibrium with the reservoir at $t=\tau$, $\rho_\tau = \rho_\tau^\text{eq}= \rho_f$---the situation usually considered in thermodynamics \cite{cen01}---then
\begin{equation}
\label{10}
 \la \Sigma\ra_\text{driving}=-\int_0^\tau dt\,\tr{\rho_t\, \pd_t \ln{\rho}_t^\text{eq}}\ .
\end{equation}
Since initial and final temperatures are the same, the mean quantum entropy production is here simply given by the irreversible work divided by the temperature,
\begin{equation}
\label{12}
  \la \Sigma\ra = \beta (\la W\ra - \Delta F) = \beta W_\text{ir} \ ,
\end{equation}
with $\Delta F = \Delta U - T \Delta S$ when $T=\text{constant}$.

c. On the other hand, for general driving, the system will be not be in equilibrium with the reservoir at the end of the driving protocol, $\rho_f\neq\rho_\tau^\text{eq}$, and an additional relaxation process will hence take place. As a result,
\begin{equation}
\label{11}
 \la \Sigma\ra_\text{driving+equil.}=-S\left(\rho_f||\rho^\text{eq}_\tau\right)-\int_0^\tau dt\,\tr{\rho_t\, \pd_t \ln{\rho}_t^\text{eq}}\ .
\end{equation}
We can therefore conclude that  the general form of the  entropy production  \eqref{7} accounts for all three kinds of irreversibilities that can occur in the driven open quantum system: initial relaxation, far from equilibrium driving and incomplete final relaxation. Equation \eqref{7} thus expresses the additivity of the entropy production. 

d. For closed, initially thermal quantum systems,  the  dynamics is unitary (without heat exchange) and $\pd_t \tr{\rho_t\ln\rho_t}=0$. The entropy production is thus \cite{def10},
\begin{equation}
\begin{split}
\label{13}
 \la \Sigma \ra &=-\int_0^\tau dt\,\tr{\rho_t\, \pd_t \ln{\rho_t^\text{eq}}}\\&=\int_0^\tau dt\,\pd_t S\left(\rho_t||\rho_t^\text{eq}\right)=S\left(\rho_\tau||\rho_\tau^\text{eq}\right)=\beta W_\text{ir}\ ,
\end{split}
\end{equation}
where the second line follows from the unitary evolution, as given by the von Neumann equation $\dot \rho_t = -i [H_t,\rho_t]$. Equation \eqref{13} provides a good approximation when the driving rate is much larger than the relaxation rate of the system and the coupling to the reservoir can be neglected. 
A classical version of Eq.~\eqref{13} can be found in Ref.~\cite{kaw07} (see also Ref.~\cite{hor09}). 
We note that for  classical open systems the total  mean nonequilibrium entropy production is  given by the classical counterpart of Eq.~\eqref{7}.

{\it Nonequilibrium entropy production rate.} The instantaneous entropy production rate is defined as the time derivative of the  nonequilibrium entropy production, $\sigma=\pd_t \Sigma$ \cite{leb08}. As discussed in the introduction, it contains essential information about the nonequilibrium dynamics of a system. Taking the derivative of Eq.~\eqref{7}, we obtain,
\begin{equation}
\begin{split}
\label{14}
 \sigma &= -\pd_t S\left(\rho_t||\rho_t^\text{eq}\right)-\tr{\rho_t\, \pd_t \ln{\rho_t^\text{eq}}} \ ,\\ 
&= -\tr{\left(\pd_t \rho_t\right) \ln{\rho_t}}+\tr{\left(\pd_t \rho_t\right) \ln{\rho_t^\text{eq}}}\ .
\end{split}
\end{equation}
The above expression agrees with the entropy production rate for slowly driven open quantum systems  derived by Breuer using a Markovian master equation of the Lindblad type \cite{bre03} (see also Ref.~\cite{lin83}). In the limit of undriven quantum systems, it reduces to the result $ \sigma = -\pd_t S\left(\rho_t||\rho_0^\text{eq}\right)$ previously derived by Spohn \cite{spo78}. We emphasize that Eq.~\eqref{14} is valid far from equilibrium.

{\it Quantum fluctuation theorem.} One can broadly distinguish two  strategies for deriving quantum fluctuation theorems. In the first approach, thermodynamic variables, like work, heat and entropy, are defined  along  single trajectories of the system \cite{esp06,roe04,roe08}. In the second approach, work and heat are determined by two energy measurements, one  taking place before and the other one after the time-dependent driving \cite{kur00,mon05,tal07,tal09,cam11}. We note, however, that the question of how to experimentally determine  thermodynamic quantities along quantum trajectories is still unsolved, contrary to the case of  energy measurements, for which a non-demolition scheme has been proposed in Ref.~\cite{hub08}. We will therefore follow the second approach.

We denote by $E_m$ and $E^R_\mu$ the respective eigenvalues of  system and  reservoir obtained after a joint measurement of their energies in  the initial state before the driving.  Similarly, we call the corresponding energy eigenvalues obtained after a joint measurement in the final equilibrium state   $E_n$ and $E^R_\nu$.  The latter can be determined simultaneously since the Hamiltonians of the system and of the reservoir commute. In the limit of weak coupling, the system-reservoir interaction energy may be neglected. For a single realization of the process, the heat exchanged by the system is thus given by the energy variation of the reservoir $Q=  -(E^R_\nu- E^R_\mu)$. We further introduce the energy change in the system, weighted by the different initial and final inverse temperatures, $\Delta E_{\beta_f,\beta_i} = \beta_f  E_n - \beta_i E_m$. The joint probability distribution $P(\Delta E_{\beta_f,\beta_i}, \beta Q)$ is then 
\begin{eqnarray}
\label{15}
\!\! &&\!\!\!\!\mathcal{P}\left(\Delta E_{\beta_f,\beta_i}, \beta Q\right)=\!\!\sum_{m,n,\mu,\nu} \!\delta\left(\Delta E_{\beta_f,\beta_i}-(\beta_f  E_n - \beta_i E_m)\right) \nonumber\\
\!\!&&\!\!\!\!\times\, \delta \left(\beta Q+\beta (E^R_\nu- E^R_\mu )\right) \,p(n,\nu|m,\mu)\,p_{m,\mu} \ ,
\end{eqnarray}
where  $p(n,\nu|m,\mu)$ is the total transition probability for system and reservoir to evolve from the state $(m,\mu)$ to the state $(n,\nu)$, and $p_{m,\mu}  = \exp(-\beta_i E_m -\beta E_\mu^R) /(Z_i Z_R)$ is the initial occupation probability. Equation \eqref{15} is  a two-temperature generalization of the joint probability distribution for internal energy change and heat discussed in Refs.~\cite{tal09,cam11}. We next define the entropy variation  that occurs in the system during a {\it single} realization of the driving protocol as,
\begin{equation}
\label{16}
\begin{split}
\Delta {\cal S}&=\beta_f (E_m- F_f) -\beta_i(E_n-F_i) \ ,\\
&= \Delta E_{\beta_f,\beta_i} -\beta_f F_f + \beta_i F_i \ ,
\end{split}
\end{equation}
where $F_i= -(1/\beta_i) \ln Z_i$   and $F_f = -(1/\beta_f) \ln Z_f$ are the  free energies of the system in the initial and the final states. Since $\ln Z_i = -\ln p_i^m - \beta_i E_m$ and $\ln Z_f = -\ln p_f^n - \beta_f E_n$, with $p_i^m$ and $p_f^n$ the initial and final  occupation probabilities of the system, we have,
\begin{equation}
\label{17}
\delta (\Delta E_{\beta_f,\beta_i} -(\beta_f  E_n - \beta_i E_m))= \delta(\Delta {\cal S}+ \ln p_f^n-\ln p_i^m) \ .
\end{equation}
The change of variable $\Delta E_{\beta_f,\beta_i} \rightarrow \Delta {\cal S}$ in Eq.~\eqref{15} then yields the joint  distribution for entropy and heat, 
\begin{eqnarray}
\label{18}
 & &\mathcal{P}\left(\Delta {\cal S}, \beta Q\right)=\sum_{m,n,\mu,\nu} \delta(\Delta {\cal S}+ \ln p_f^n-\ln p_i^m)\nonumber\\
& &\times\, \delta \left(\beta Q+\beta (E^R_\nu- E^R_\mu )\right) \,p(n,\nu|m,\mu)\,p_{m,\mu} \ .
\end{eqnarray}
According to Eq.~(2) the nonequilibrium entropy production for a single realization is
\begin{equation}
\label{19}
\Sigma = \Delta {\cal S} -\beta Q \ .
\end{equation}
The entropy production $\Sigma$  is thus the sum of the entropy variation of the system $\Delta {\cal S}$  and of the reservoir $ \Delta {\cal S}^R = -\beta Q$, and  hence corresponds to the total entropy change.
The  distribution of the  entropy production is obtained by integrating Eq.~\eqref{18} over the heat. We find,
\begin{eqnarray}
\label{20}
&&\mathcal{P}(\Sigma)=\int d(\beta Q)\,\mathcal{P}(\beta Q+\Sigma,\beta Q) \nonumber \\
&&= \sum_{m,n,\mu,\nu} \delta(\Sigma + \ln p_f^n-\ln p_i^m -\beta (E^R_\nu- E^R_\mu )) \nonumber \\
&& \hspace{1.3cm}\times\,p(n,\nu|m,\mu)\,p_{m,\mu} \ .
\end{eqnarray}
One can verify that the mean quantum nonequilibrium entropy production $\la \Sigma \ra$ computed directly  from Eq.~\eqref{20} leads to the general expression \eqref{7}, as it should. To finally establish the quantum fluctuation theorem, we  evaluate,
\begin{eqnarray}
\label{21}
\la \e{-\Sigma}\ra&=&\int d\Sigma  \, \e{-\Sigma} \mathcal{P}(\Sigma) \nonumber \\
&=&\sum_{m,n,\mu,\nu}p(n,\nu|m,\mu)\,p_{n,\nu} = 1 \  ,
\end{eqnarray}
with $p_{n,\nu}  = \exp(-\beta_f E_n -\beta E_\nu^R) /(Z_f Z_R)$. Here, we have used that the reservoir remains in equilibrium at all times. The last equality follows from the unitary dynamics of system-plus-reservoir and the normalization of the corresponding total density operator. 
Equation \eqref{21}  is valid for any initial equilibrium condition of the  quantum system---\mbox{not necessarily  equilibrium with the reservoir.} 

 So far, we have restricted ourselves to  initial and final equilibrium states for which the thermodynamic entropy is  well-defined. However,  by twice measuring  the  density operator of the system, instead  of its Hamiltonian, and obtaining the  two states $|r \rangle $ and $ |s \rangle $, we can  introduce the information entropy production for a single realization,
\begin{equation}  
\Sigma_I =  - \ln \la s|\rho_f |s\ra + \ln  \la r|\rho_i |r\ra -\beta (E^R_\nu- E^R_\mu) \ .
\end{equation}
After averaging, the first two terms on the right-hand side become the von Neumann (information) entropy of  final and initial states. An analogous derivation  then shows that the quantity $\Sigma_I$ satisfies a fluctuation theorem, $\la \exp (-\Sigma_I)\ra =1$, that is valid for any initial and final nonequilibrium states. The latter is the quantum generalization of the integral fluctuation theorem proposed by  Seifert \cite{sei05}. The corresponding mean information entropy production is  given by Eq.~\eqref{7} with arbitrary $\rho_i$ and $\rho_f$. In particular, for an undriven partial relaxation process from a nonequilibrium  state to another, we have,
\begin{equation} 
\la \Sigma_I\ra=S\left(\rho_i||\rho_0^\text{eq}\right)-S\left(\rho_f||\rho^\text{eq}_\tau\right) \ ,
\end{equation}
which is consistent with the result  obtained in Ref.~\cite{esp10},  in the limit of weak system-reservoir coupling.

{\it Conclusion.} We have developed a generic  formalism which allows  the determination of  the exact mean nonequilibrium entropy production, and entropy production rate, for open quantum systems  driven arbitrarily far from equilibrium. 
Moreover, by using a joint system-reservoir-measurement approach, we have derived a quantum extension of the integral fluctuation theorem of Seifert for the total entropy change. 

This work was supported by  the Emmy Noether Program of the DFG (contract No LU1382/1-1) and the cluster of excellence Nanosystems Initiative Munich (NIM). 


\begin{thebibliography}{99}
\bibitem{rem} The case of several heat (or particle) reservoirs can be treated in a similar fashion.
\end{thebibliography}
\end{document}